\documentclass{PoS}
\usepackage{graphicx}

\title{Heavy neutrino potential \\ for neutrinoless double beta decay}

\ShortTitle{Heavy neutrino potential ...}

\author{Yoritaka Iwata$^{1,2}$
\thanks{The author would like to express his sincere gratitude to Dr. J. Men\'endez for fruitful comments.Numerical calculations were carried out at the workstation system of Institute of Innovative Research, Tokyo Institute of Technology, and COMA of University of Tsukuba..}\\
$^{1}$Institute of Innovative Research, Tokyo Institute of Technology, Japan \\
$^{2}$Department of Mathematics, Shibaura Institute of Technology, Japan\\
        E-mail: \email{iwata\_phys@08.alumni.u-tokyo.ac.jp}}

\abstract{
Heavy neutrino potential for neutrinoless double beta decay is studied with focusing on its statistical property.
The existence condition for heavy neutrinos is also presented.
In particular sterile neutrinos are possible candidate for heavy nuetrinos.
The statistics provide a gross view of understanding amplitude of constitutional components of the nuclear matrix element.
}

\FullConference{The 26th International Nuclear Physics Conference\\
		11-16 September, 2016\\
		Adelaide, Australia}

\begin{document}
\section{Introduction}
Observation of neutrino oscillation has clarified the nonzero neutrino mass.
The observation of neutrinoless double-beta decay, for whose existence nonzero neutrino mass plays a supportive role, is associated with important physics; e.g., 
\begin{itemize}
\item existence of Majorana particle,  
\item breaking of leptonic number conservation,  
\item quantitative determination of neutrino mass.
\end{itemize}
In this sense neutrinoless double-beta decay is intriguing enough to bring about an example exhibiting the physics beyond the standard model of elementary particle physics (for a review, see \cite{16engel}).
Although LSND~\cite{06dodelson} experiment has suggested the possible existence of heavy neutrinos (recognized as ``sterile neutrino'' in various literatures), theorists started to account for such a contribution to the neutrinoless double beta decay half life only recently.
In addition, regarding the motivation for sterile neutrinos, the GALLEX/SAGE experiments~\cite{97bahcall} and the reactor anomaly support such existences. 
All three experiments suggest neutrino masses on the eV scale. 
Another motivation is that sterile neutrinos could be dark matter candidates, in that case the masses are on the keV scale.

If heavy neutrinos exist, those neutrinos are mixed into the effective mass.
As an example of relation between the half life of neutrinoless double-beta decay, the effective light neutrino mass ($m_{\nu}$), and  the effective heavy neutrino mass ($\eta_{N}$) is given by \cite{12vergados,13horoi}
\begin{equation} \label{halflife} \begin{array}{ll}
[T_{0 \nu}^{1/2}] ^{-1} = G \left\{ |M^{0 \nu}|^2\left ({\displaystyle \frac{m_{\nu}}{m_e}} \right)^2 +  |M^{0 N}|^2\left (\eta_N \right)^2 \right\}, 
\end{array}  \end{equation}
where $G$ is the phase space factor (its value is obtained rather precisely), $m_e$ is the electron mass (its value is also precisely obtained), $\eta_{N}$ denotes the effective mass relative to electron mass, and $M^{0 \nu}$ and $M^{0 N}$ are the nuclear matrix elements (NME, for short) for light and heavy neutrinos respectively. 
In this context light neutrinos mean already-observed ordinary neutrinos.
Under the existence of heavy neutrino, we need to have half life observations for two different double-beta decay events (for example, decay of calcium and xenon):
\begin{equation} \label{halflife21} \begin{array}{ll}
[T_{0 \nu,{\rm I}}^{1/2}] ^{-1} =  G_I \left\{ |M^{0 \nu}_{\rm I}|^2 \left ({\displaystyle \frac{m_{\nu}}{m_e}} \right)^2 +  |M_{\rm I}^{0 N}|^2\left (\eta_N \right)^2 \right\},
\end{array}  \end{equation}
and
\begin{equation} \label{halflife22} \begin{array}{ll} 
[T_{0 \nu,{\rm II}}^{1/2}] ^{-1} =  G_{II} \left\{ |M^{0 \nu}_{\rm II}|^2 \left ({\displaystyle \frac{m_{\nu}}{m_e}} \right)^2 +  |M_{\rm II}^{0 N}|^2\left (\eta_N \right)^2 \right\},
\end{array}  \end{equation}
where indices ${\rm I}$ and ${\rm II}$ identify the kind of decaying nuclei. 
Because there are two unknown quantities: $m_{\nu}$ and $\eta_N$, here we have two equations. 
In order to determine the neutrino mass, it is necessary to calculate $M_{\rm I}^{0 \nu}$, $M_{\rm II}^{0 \nu}$, $M_{\rm I}^{0 N}$ and $M_{\rm II}^{0 N}$ very precisely. 
At this point, many calculations by various theoretical models have been dedicated to NME calculations.   
Since the detail information on initial and final states (i.e., quantum level structure of these states) is necessary for the calculation of NMEs, it is impossible to have reliable NME without knowing nuclear structures. 
The impact of precise NME calculations is expected to be large enough (e.g., for a large-scale shell model calculation for light neutrinos, see Ref.~\cite{15iwata}), and the unknown leptonic mass-hierarchy and the Majorana nature of neutrinos are expected to be discovered. 

In this article heavy neutrino potential for neutrinoless double beta decay (for the definition, see Eq.~(\ref{nupot})) is studied from a statistical point of view.
The results in this article are intended to be compared to light neutrino cases (that is, ordinary neutrino case) presented in Ref.~\cite{16iwata}.
The comparison clarifies the contribution of heavy neutrinos for neutrinoless double beta decay half-life.

\section{Condition for the existence of heavy neutrino}
Role of the nuclear matrix element is seen by solving Eqs.~(\ref{halflife21})-(\ref{halflife22}).
Under the validity of 
\begin{equation}  \label{cond} \begin{array}{ll}
 |M_{\rm I}^{0 N}|^2 /  |M^{0 \nu}_{\rm I}|^2  \ne |M_{\rm II}^{0 N}|^2 /  |M^{0 \nu}_{\rm II}|^2,  
\end{array}  \end{equation} 
the effective neutrino mass for light and heavy neutrinos are represented by
\begin{equation} \label{cond1} \begin{array}{ll}
\left ({\displaystyle \frac{m_{\nu}}{m_e}} \right)^2 = 
{\displaystyle \frac{ - |M^{0 N}_{\rm II}|^2 [G_{I} T_{0 \nu,{\rm I}}^{1/2}] ^{-1} + |M^{0 N}_{\rm I}|^2 [G_{II} T_{0 \nu,{\rm II}}^{1/2}] ^{-1} } {  |M^{0 \nu}_{\rm II}|^2  |M_{\rm I}^{0 N}|^2 - |M^{0 \nu}_{\rm I}|^2  |M_{\rm II}^{0 N}|^2 } },  
\end{array}  \end{equation}
and
\begin{equation}  \label{cond2} \begin{array}{ll}
\left (\eta_N \right)^2 = 
{\displaystyle \frac{ |M^{0 \nu}_{\rm II}|^2 [G_{I} T_{0 \nu,{\rm I}}^{1/2}] ^{-1} - |M^{0 \nu}_{\rm I}|^2 [G_{II} T_{0 \nu,{\rm II}}^{1/2}] ^{-1} } {  |M^{0 \nu}_{\rm II}|^2  |M_{\rm I}^{0 N}|^2 - |M^{0 \nu}_{\rm I}|^2  |M_{\rm II}^{0 N}|^2 } },
\end{array}  \end{equation}
respectively.
The condition (\ref{cond}) is valid if nuclear structure effect on double beta decay is not so simple; indeed it is not true only if NMEs for different decay candidates are exactly the same in their heavy-to-light ratios.
This condition was explored in Refs. \cite{11faessler,15lisi}.
According to Eq.~(\ref{cond1}),  the experimentally-confirmed nonzero neutrino effective mass suggests that
\begin{equation} \label{determ00} \begin{array}{ll}
 |M^{0 N}_{\rm I}   |^2 G_{I} T_{0 \nu,{\rm I}}^{1/2}   \ne {\displaystyle  |M^{0 N}_{\rm II}   |^2 G_{II} T_{0 \nu,{\rm II}}^{1/2} } 
\end{array}  \end{equation}
where note that Eq.~(\ref{determ00}) is written only by heavy neutrino NMEs and half lives.
According to Eq.~ (\ref{cond2}),
\begin{equation} \label{determ01}
 |M^{0 \nu}_{\rm I}|^2   G_{I} T_{0 \nu,{\rm I}}^{1/2} = { |M^{0 \nu}_{\rm II}|^2   G_{II} T_{0 \nu,{\rm II}}^{1/2} }  
\end{equation}
suggests that heavy neutrinos do not exist. 
The satisfaction of Eq.~(\ref{determ01}) means either one of the following possibilities:  \vspace{0.6mm} 

 (i) the present framework~(\ref{halflife}) is too simple to be valid, \vspace{1.2mm} 
 
 (ii) heavy neutrinos do not exist.  \vspace{1.2mm} \\
Since Eq.~(\ref{determ01}) is written even without knowing anything about heavy neutrino, this condition is practically used as the sufficient condition for the existence of heavy neutrino under the validity of the framework~(\ref{halflife})  (i.e., heavy neutrino existence condition for Eq.~(\ref{halflife})).
It is worth noting that, as discussed around Eq.~(15) of Ref.~\cite{16engel}, additional terms can be added to Eq.~(\ref{halflife}). 
Under the non-existence of heavy neutrino (by applying Eq.~(\ref{determ01}) to Eq.~(\ref{cond1})),
\begin{equation} \begin{array}{ll}
\left ({\displaystyle \frac{m_{\nu}}{m_e}} \right)^2 =  {\displaystyle \frac{ [G_{I} T_{0 \nu,{\rm I}}^{1/2}] ^{-1}}{ |M^{0 \nu}_{\rm I}|^2} }
{\displaystyle \frac{  |M^{0 \nu}_{\rm II}|^2  |M^{0 N}_{\rm I}|^2    -  |M^{0 \nu}_{\rm I}|^2  |M^{0 N}_{\rm II}|^2 }
 {  |M^{0 \nu}_{\rm II}|^2  |M_{\rm I}^{0 N}|^2 - |M^{0 \nu}_{\rm I}|^2  |M_{\rm II}^{0 N}|^2 } }  
 =  {\displaystyle \frac{1}{G_I}} {\displaystyle \frac{ [T_{0 \nu,{\rm I}}^{1/2}] ^{-1}}{ |M^{0 \nu}_{\rm I}|^2} }
\end{array}  \end{equation}
trivially follows.
If the squared masses are positive,
\begin{equation} \begin{array}{ll} 
\left( {\displaystyle |M^{0 N}_{\rm II}|^2 [G_{I} T_{0 \nu,{\rm I}}^{1/2}] ^{-1} - |M^{0 N}_{\rm I}|^2 [G_{II} T_{0 \nu,{\rm II}}^{1/2}] ^{-1} } \right)
\left( {\displaystyle |M^{0 \nu}_{\rm II}|^2 [G_{I} T_{0 \nu,{\rm I}}^{1/2}] ^{-1} - |M^{0 \nu}_{\rm I}|^2 [G_{II} T_{0 \nu,{\rm II}}^{1/2}] ^{-1} } \right) \le 0, 
\end{array}  \end{equation}
must be satisfied (i.e., real mass condition).

\section{Neutrino potential}
\subsection{Nuclear matrix element} 
Nuclear matrix element in double beta decay is investigated under the closure approximation.
It approximates all the different virtual intermediate energies by a single intermediate energy (i.e., with the averaged energy called closure parameter).
For neutrinoless double beta decay, nuclear matrix element for light and heavy neutrinos are written by 
\begin{equation} \label{matrixel} \begin{array}{ll}
M^{0 \nu} = M_{\rm F}^{0 \nu} - \frac{g_V^2}{g_A^2} M_{\rm GT}^{0 \nu} + M_{\rm T}^{0 \nu} 
\end{array} \end{equation}
and
\begin{equation} \label{matrixel} \begin{array}{ll}
M^{0 N} = M_{\rm F}^{0 N} - \frac{g_V^2}{g_A^2} M_{\rm GT}^{0 N} + M_{\rm T}^{0 N} \\
\end{array} \end{equation}
respectively, where $g_V$ and $g_A$ denote vector and axial coupling constants, and $\alpha$ of $M_{\alpha}^{0 \nu}$ is the index for the double beta decay of three kinds: $\alpha =$ F, GT, T (Fermi, Gamow-Teller, and tensor parts). 
According to Ref.~\cite{iwata-cns},  each part is further represented by the sum of two-body transition density (TBTD) and anti-symmetrized two-body matrix elements. 
\begin{equation} \label{me-general} \begin{array}{ll}
M_{\alpha}^{0 x}  = \langle 0_f^+ |O_{\alpha}^{0 x} | 0_i^+ \rangle \vspace{2.5mm} \\
= \sum {\rm TBTD}(n'_1 l'_1 j'_1 t'_1, n'_2 l'_2 j'_2 t'_2, n_1 l_1 j_1 t_1,  n_2 l_2 j_2 t_2; J)  \vspace{1.5mm} \\
\quad \langle n'_1 l'_1 j'_1 t'_1, n'_2 l'_2 j'_2 t'_2; J|O_{\alpha}^{0 x}(r) | n_1 l_1 j_1 t_1, n_2 l_2 j_2 t_2; J  \rangle_{\rm AS}
\end{array} \end{equation}
where $O^{0 x}_{\alpha}(r)$ are transition operators of neutrinoless double beta decay, and $0_i^+$ and $0_f^+$ denote initial and final states, respectively ($x$ is either $\nu$ or $N$).
The sum is taken over indices $( n_i l_i j_i t_i,n'_j l'_j j'_j t'_j)$ with $i,j=1,2$, where $n$, $l$, $j$ and $t$ mean principal, angular momentum and isospin quantum numbers, respectively, $j_1$ and $j_2$ (or $j_1'$ and $j_2'$) are coupled to $J$ (or $J$), similarly $l_1$ and $l_2$ (or $l_1'$ and $l_2'$) are coupled to $\lambda$ (or $\lambda'$), and $t_1 = t_2 = 1/2, ~ t_1' = t_2' = -1/2$ is valid if neutrons decay into protons.

\begin{figure*} [t]
\includegraphics[width=156mm]{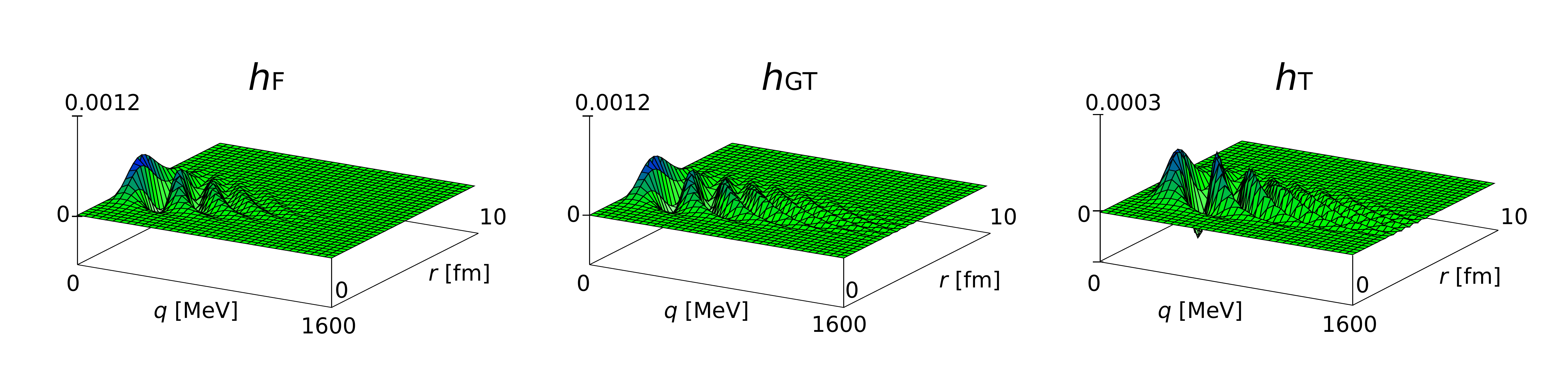}  \\
(a) $n=n' =0$ and $l=l'=3$ \\
\includegraphics[width=156mm]{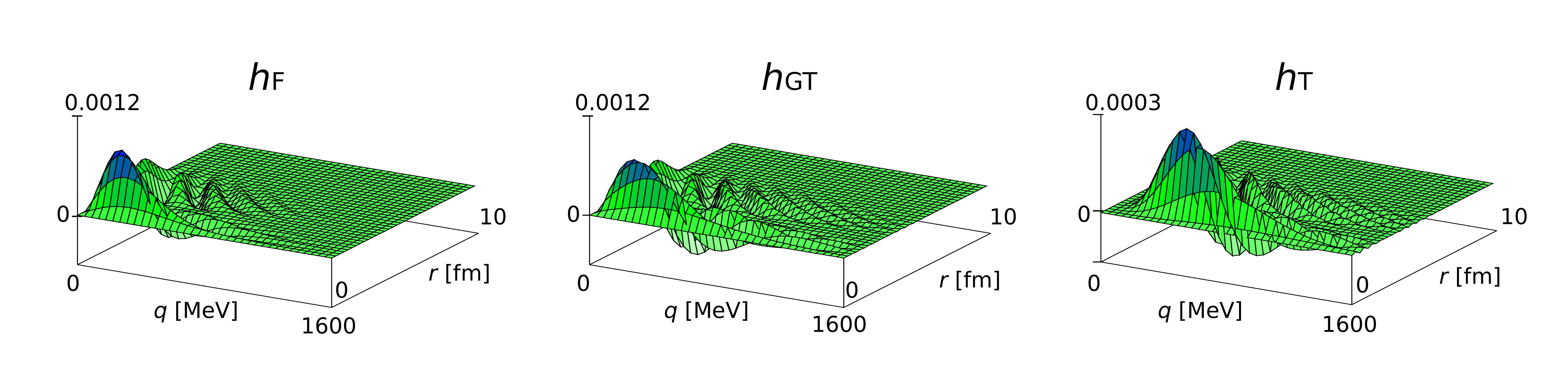}  \\
(b) $n=n' =1$ and $l=l'=0$ \\
\caption{ \label{fig1} (Color online)
Integrands of Eq.~({\protect \ref{nupot4}}) are depicted for $n=n' =0$ and $l=l'=3$ in panel a, and for $n=n' =1$ and $l=l'=0$ in panel b.
The plots are made for $r=\sqrt{2}\rho=0$ to 10~fm and $q=0$ to 1600~MeV.
The closure parameter $\langle E \rangle$ is fixed to 0.5~MeV, which is suggested by the calculation without using closure approximation~{\protect \cite{13senkov}}.
}
\end{figure*}

The two-body matrix element before the anti-symmetrization is represented by
\begin{equation} \begin{array}{ll}
\langle n'_1 l'_1 j'_1 t'_1, n'_2 l'_2 j'_2 t'_2; J|O_{\alpha}^{0 x}(r) | n_1 l_1 j_1 t_1, n_2 l_2 j_2 t_2; J  \rangle  \vspace{2.5mm} \\
 = 2 {\displaystyle \sum_{S, S', \lambda, \lambda'}} \sqrt{j_1' j_2' S' \lambda'} \sqrt{j_1 j_2 S \lambda}  
~ \langle l_1' l_2' \lambda' S'; J| S_{\alpha} |  l_1 l_2 \lambda S; J \rangle 
~ \langle n_1' l_1' n_2' l_2'; J| H_{\alpha}(r) |  n_1 l_1 n_2 l_2 \rangle  \vspace{1.5mm} \\
\quad \left\{
\begin{array} {ccc}
l_1' & 1/2 & j_1' \\
l_2' & 1/2 & j_2' \\
\lambda' & S' & J 
\end{array}
\right\} 
~ \left\{
\begin{array} {ccc}
l_1 & 1/2 & j_1 \\
l_2 & 1/2 & j_2 \\
\lambda & S & J 
\end{array}
\right\}
\end{array} 
\end{equation}
where $H_{\alpha}(r)$ is the neutrino potential,  $S_{\alpha}$ denotes spin operators, $S$ and $S'$ mean the two-body spins, and $\{ \cdot \}$ including nine numbers denotes the 9j-symbol.
By implementing the Talmi-Moshinsky transforms:
\begin{equation}
 \langle n l, NL|  n_1 l_1, n_2 l_2 \rangle_{\lambda} 
 \langle n' l', N'L'|  n_1' l_1', n_2' l_2' \rangle_{\lambda'}  
\end{equation}
 the harmonic oscillator basis is transformed to the center-of-mass system.
\begin{equation} \begin{array}{ll}
 \langle l_1' l_2' \lambda' S'; J| S_{\alpha} |  l_1 l_2 \lambda S; J \rangle 
 \langle n_1' l_1' n_2' l_2'; J| H_{\alpha}(r) |  n_1 l_1 n_2 l_2 \rangle \vspace{3.5mm} \\
= {\displaystyle \sum_{\rm mos2}} \langle n l, NL|  n_1 l_1, n_2 l_2 \rangle_{\lambda} 
 \langle n' l', N'L'|  n_1' l_1', n_2' l_2' \rangle_{\lambda'}  
 \langle l' L \lambda' S'; J|  S_{\alpha} | l L \lambda S; J \rangle 
 \langle n' l'|  H_{\alpha}(\sqrt{2} \rho) | n l \rangle, 
\end{array}  \end{equation}
where $\rho = r /\sqrt{2}$ is the transformed coordinate of center-of-mass system, and ``mos2'' means that the sum is taken over $(n,n',l,l',N,N')$~\cite{iwata-cns}.
In this article, in order to have a comparison to the preceding results \cite{16iwata}, we focus on the neutrino potential effect arising from
\begin{equation} \label{nupot} \begin{array}{ll}
 \langle n' l'|  H_{\alpha}(\sqrt{2} \rho) | n l \rangle.
\end{array}  \end{equation}
This part is responsible for the amplitude of each transition from a state with $n$, $l$ to another state with $n'$, $l'$, while the cancellation is determined by spin-dependent part.
For calculations of heavy-neutrino exchange matrix elements, see Refs.~\cite{10blennow,14faessler,15barea,15hyvarinen,16horoi}.

\begin{figure*} [t]
\includegraphics[width=150mm]{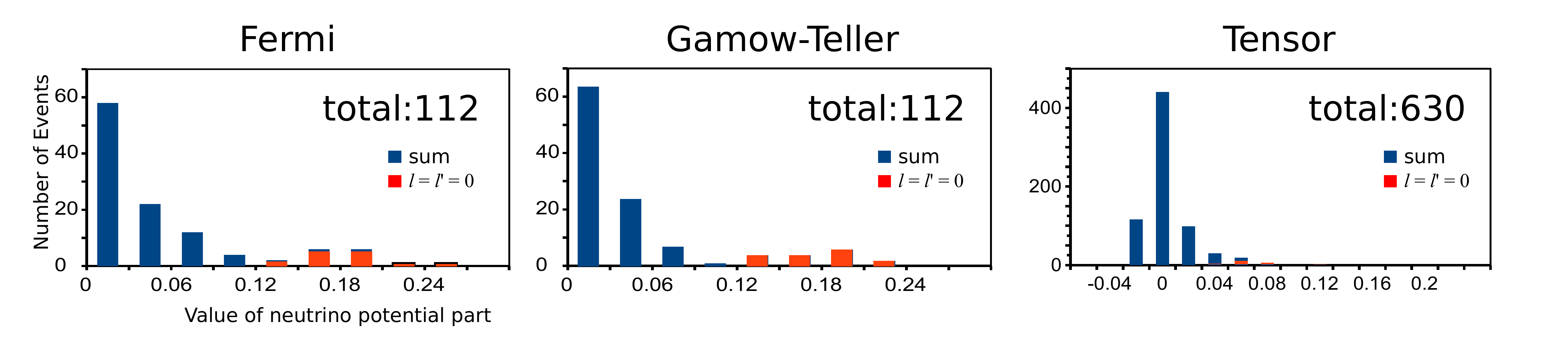} 
\caption{ \label{fig2} (Color online)
Frequency distribution of $\langle n' l'|  H_{\alpha}(\sqrt{2} \rho) | n l \rangle$ is shown limited to nonzero cases.
Cases with $n, n' = 0, 1, \cdots, 3$ and  $l, l' = 0, 1, \cdots, 6$ are taken into account, where note that $l\ne l'$ results in $\langle n' l'|  H_{\alpha}(\sqrt{2} \rho) | n l \rangle = 0$ in Fermi and Gamow-Teller cases~{\protect \cite{iwata-cns}}.
The total number of events with nonzero $\langle n' l'|  H_{\alpha}(\sqrt{2} \rho) | n l \rangle$ is shown in each panel.
}
\end{figure*}

\begin{table*}[t]
\caption{Large contributions are listed from 1st to 10th largest ones. 
Two symmetric cases resulting in an equivalent value are shown in the same position. }
\begin{center}
  \begin{tabular*}{\columnwidth}{@{\extracolsep{\fill}}c | cc| c c| c c} \hline \hline
 &  \multicolumn{2}{c|}{Fermi} &  \multicolumn{2}{c|}{Gamow-Teller} &  \multicolumn{2}{c}{Tensor} \\ 
Ranking &  $(n~ l~ n'~ l')$  & Value &  $(n~ l~ n'~ l')$  & Value & $(n~ l~ n'~ l')$ & Value  \\ \hline
 1   & (0~0~0~0) &  $0.261$ &  (0~0~0~0)  &  $0.230$ &  (0~0~0~0) & 0.125  \\
 2  &  (1~0~1~0) & $0.232$ &  (1~0~1~0) &  $0.217$ &  (1~0~1~0) &  $0.099$  \\
 3   & (2~0~2~0) & $0.210$  &  (2~0~2~0)  & $0.207$ &  (0~0~1~0) & $0.088$  \\
      &   &  &   &   & (1~0~0~0)  &  \\ 
 4   & (0~0~1~0) & $0.193$  &  (3~0~3~0)  & $0.198$ &  (2~0~2~0) & $0.083$  \\
     &  (1~0~0~0)  &  &   &   &  &  \\
 5   & (3~0~3~0) & $0.192$  &  (1~0~2~0)  & $0.1809$ &  (1~0~2~0) & $0.080$  \\  
     &  &  &   (2~0~1~0)  &   & (2~0~1~0)  &  \\
 6   & (1~0~2~0) & $0.190$  &  (2~0~3~0)  & $0.1806$ &  (0~0~0~1) & $0.072$  \\  
      &  (2~0~1~0) &  &  (3~0~2~0)   &   &  (0~1~0~0)  &    \\
  7   & (2~0~3~0) & $0.179$  &  (0~0~1~0)  & $0.171$ &  (3~0~3~0) & $0.0714$  \\ 
      &  (3~0~2~0) &  & (1~0~0~0)  &   &  &    \\ 
  8   & (1~0~3~0) & $0.161$  &  (1~0~3~0)  & $0.160$ &  (0~0~1~1) & $0.0710$  \\  
      &  (3~0~1~0) &  &  (3~0~1~0)    &   &   (1~1~0~0) &    \\ 
  9   & (0~0~2~0) & $0.157$  &  (0~0~2~0)  & $0.145$ &  (2~0~3~0) & $0.070$  \\
   &  (2~0~0~0) &  &  (2~0~0~0)   &   &   (3~0~2~0)  &   \\    
 10   & (0~0~3~0) & $0.132$  &  (0~0~3~0)  & $0.130$ &  (0~0~2~0) & $0.065$  \\
   & (3~0~0~0)   &  &  (3~0~0~0)  &   &  (2~0~0~0)  &    \\    
\hline  \hline
 \end{tabular*}
\end{center}
\label{table1}
\end{table*}

\subsection{Neutrino potential represented in the center-of-mass system}
Under the closure approximation neutrino potential~\cite{91tomoda,10horoi, 13senkov} is represented by
\begin{equation} \label{nupot3} \begin{array}{ll}
H_{\alpha}(\sqrt{2} \rho) = \frac{2R}{\pi} {\displaystyle \int_0^{\infty} } f_{\alpha} (\sqrt{2} \rho q) \frac{h_{\alpha}(q)}{  \sqrt{q^2 + m_{\nu}^2}  ( \sqrt{q^2 + m_{\nu}^2}  + \langle E \rangle ) } ~q^2 ~ dq.
\end{array} \end{equation}
where $q$ is the momentum of virtual neutrino, $m_{\nu}$ is the effective neutrino mass, $R$ denotes the radius of decaying nucleus, and $f_{\alpha}$ is a spherical Bessel function ($\alpha=0,2$), 
In particular $\langle E \rangle$ is called the closure parameter, which means the averaged excitation energy of virtual intermediate state.
In Eq.~(\ref{nupot3}) neutrino potentials include the dipole form factors (not just the form factors) that take into account the nucleon size.
The massless neutrino limit ($m_{\nu} \to 0$) of neutrino potential is
\begin{equation} \label{nupot2} \begin{array}{ll}
H_{\alpha}(\sqrt{2} \rho) = \frac{2R}{\pi} {\displaystyle \int_0^{\infty} } f_{\alpha} (\sqrt{2} \rho q) \frac{h_{\alpha}(q)}{q+ \langle E \rangle} ~q ~ dq, 
\end{array} \end{equation}
and the heavy mass limit ($m_{\nu} >>  \langle E \rangle$, ~$m_{\nu}^2 >> q^2$) of neutrino potential is 
\begin{equation} \label{nupot4} \begin{array}{ll}
H_{\alpha}(\sqrt{2} \rho) = \frac{1}{m_{\nu}^2} \frac{2R}{\pi} {\displaystyle \int_0^{\infty}} f_{\alpha} (\sqrt{2} \rho q) h_{\alpha}(q) ~q^2 ~ dq, 
\end{array} \end{equation}
For ordinary light neutrinos, the neutrino potential in the massless limit can be utilized. 
Simkovic unit is exploited for heavy neutrino case, in which the value of $m_{\nu}^2 H_{\alpha}(\sqrt{2} \rho)$ is divided by proton and electon masses (i.e. the value of $(m_{\nu}^2/m_p m_e)H_{\alpha}(\sqrt{2} \rho)$ is shown in this article).
Following the corresponding study on massless limit cases \cite{16iwata}, this article is devoted to investigate heavy mass limit cases.

The representation of neutrino potentials are
\begin{equation} \label{eq-form} \begin{array}{ll}
h_{\rm F}(q^2) = \frac{g_V^2}{(1+q^2/\Lambda_V^2)^4}  \vspace{2.5mm} \\
h_{\rm GT}(q^2) =  \frac{2}{3} \frac{q^2}{4 m_p^2} (\mu_p - \mu_n) ^2 \frac{g_V^2}{(1+q^2/\Lambda_V^2)^4}  
 +
 \left( 1-\frac{2}{3} \frac{q^2}{q^2+m_{\pi}^2} + \frac{1}{3} \left( \frac{q^2}{q^2+m_{\pi}^2} \right)^2 \right)
 \frac{g_A^2}{(1+q^2/\Lambda_A^2)^4} 
 \vspace{2.5mm} \\
h_{\rm T}(q^2) =  \frac{1}{3} \frac{q^2}{4 m_p^2} (\mu_p - \mu_n) ^2 \frac{g_V^2}{(1+q^2/\Lambda_V^2)^4}  
 +
\left( \frac{2}{3} \frac{q^2}{q^2+m_{\pi}^2} - \frac{1}{3} \left( \frac{q^2}{q^2+m_{\pi}^2} \right)^2 \right) 
 \frac{g_A^2}{(1+q^2/\Lambda_A^2)^4} 
 \vspace{2.5mm} \\
\end{array} \end{equation}
where $\mu_p$ and $\mu_n$ are magnetic moments satisfying $\mu_p - \mu_n = 4.7$, $m_p$ and $m_{\pi}$ are proton mass and pion mass, and $\Lambda_V=850$MeV, $\Lambda_A=1086$MeV are the finite size parameters.

\begin{figure*} [t]
\includegraphics[width=145mm]{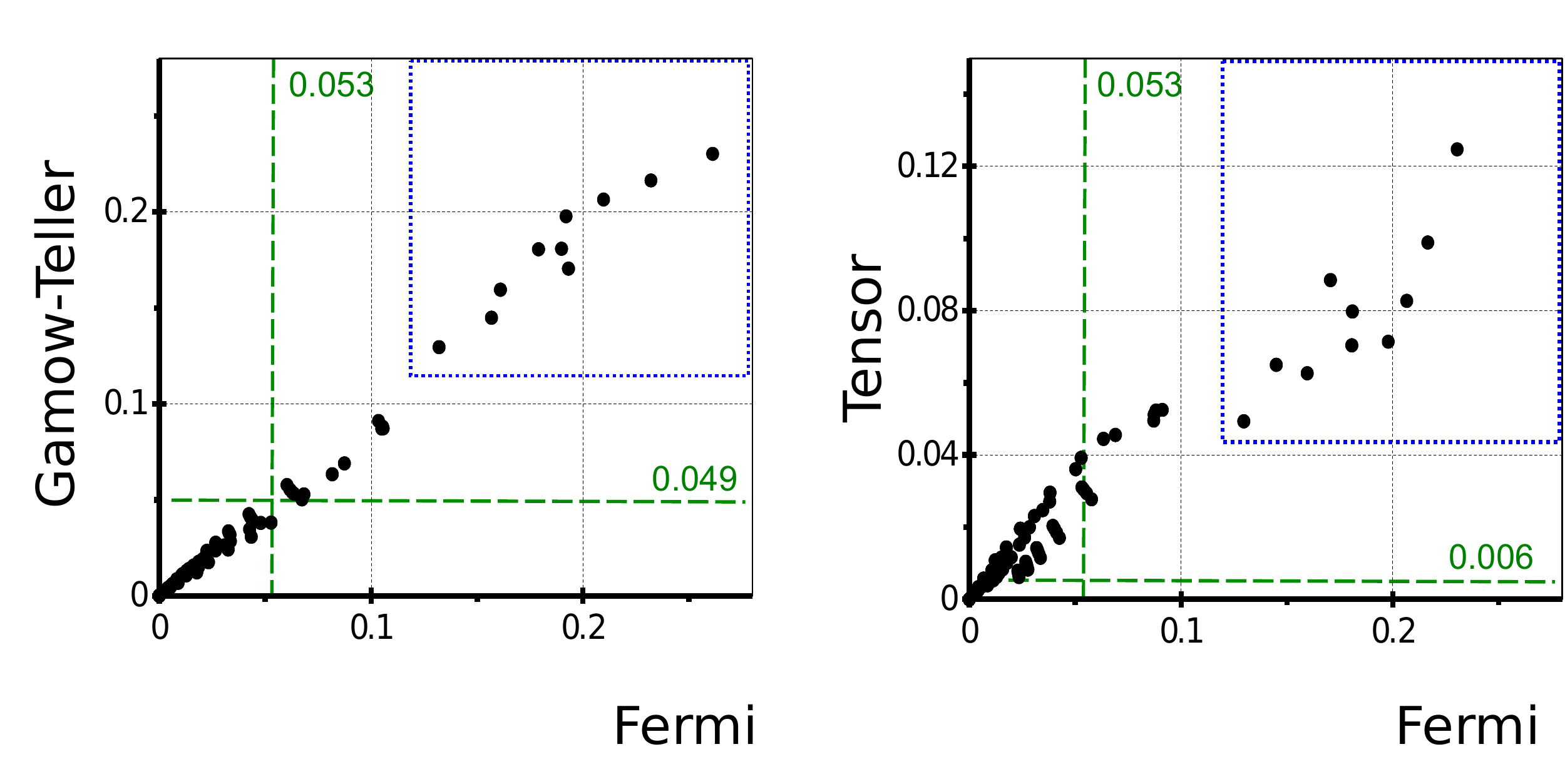} 
\caption{ \label{fig3} (Color online)
Correlation between Eq.~({\protect \ref{nupot4}}) values are examined by assuming $l=l'$.
[Left] Correlation between Eq.~({\protect \ref{nupot4}}) values for Fermi and Gamow-Teller parts, where the condition $l=l'$ does not bring about any limitations for Fermi and Gamow-Teller parts. 
[Right] Correlation between Eq.~({\protect \ref{nupot4}}) values for Fermi and tensor parts, where values for the tensor part is always positive if $l=l'$ is assumed. 
For both panels,  all the top 10 contributions listed in Table~{\protect \ref{table1}} are included in dotted-blue rectangles, and the average of all the nonzero contributions are shown in green dashed lines.
}
\end{figure*}

Figure~\ref{fig1} shows the integrand of Eq.~(\ref{nupot4}).
In any case ripples of the form: $q \rho$ = const. can be found if $q$ and $\rho$ are relatively large.
The upper-value of the integral range should be at least equal to or larger than $q=1600$~MeV.  
In our research including our recent publication~\cite{15iwata}, we take $q=2000$~MeV and $r=10$~fm as the maximum value for numerical integration of Eq.~(\ref{nupot4}) (massless neutrino cases).
We noticed that, for the convergence, $q_{\max}$ of the integral should be rather larger for the heavy cases compared to the light cases. 

\section{Statistics}
Since actual quantum states are represented by the superposition of basic states such as $| nl \rangle$ in the shell-model treatment, the contribution of neutrino potential part can be regarded as the superposition:
\begin{equation} \begin{array}{ll}
 {\displaystyle \sum_{n, n', l, l'}}  k_{n, n', l, l'} ~ \langle n' l'|  H_{\alpha}(\sqrt{2} \rho) | n l \rangle.
\end{array} \end{equation}
using a suitable set of coefficients $\{ k_{n, n', l, l'} \}$ determined by the nuclear structure of grandmother and daughter nuclei.
Accordingly, in order to see the difference between the light and heavy neutrino contributions, it is worth investigating the statistical property of neutrino potential part~(\ref{nupot}) at heavy mass limit. 

Frequency distribution of neutrino potential part~(\ref{nupot}) is shown in Fig.~\ref{fig2}.
The values are always positive for Fermi and Gamow-Teller parts, while the tensor part includes non-negligible negative values.
Indeed, the sum of positive and negative contributions of tensor part suggests that total sum 9.458 is obtained by the cancellation between $+9.943$ and $-0.485$ (i.e., $9.458 = 9.943-0.485$).
The order of the magnitude is different only for the tensor part.
Indeed, the average of the nonzero components is 0.0526 for the Fermi part, 0.0485 for the Gamow-Teller part, and 0.0063 for the tensor part.
Contributions with $l= l' =0$ (sum) cover 49.6$\%$ of the total contributions (sum) for Fermi, 52.3$\%$ for Gamow-Teller parts, and 12.8$\%$ for tensor part.
Since the corresponding values in light ordinary neutrino cases are 27.1$\%$ for Fermi part,  27.1$\%$ for Gamow-Teller part, and 7.2$\%$ for tensor part \cite{16iwata}, $l= l' =0$ component is clarified to play a more dominant role (roughly equal to twice) in heavy neutrino case.

Large contributions for Fermi, Gamow-Teller and tensor parts are summarized in Table~\ref{table1}.
Contribution labeled by $(n~ l~ n'~ l')=(0~0~0~0)$ (i.e. transition between $0s$ orbits) provides the largest contribution in any part. Roughly speaking, we see that $s$-orbit is remarkably significant in heavy neutrino cases.
Indeed, all the top 10 contributions of Fermi and Gamow-Teller parts are completely filled with $s$-orbit contributions.
As seen in the top 10 list the order of the kind $(n~ l~ n'~ l')$ are similar for Fermi and Gamow-Teller parts, where note that the order of Fermi and Gamow-Teller parts is exactly the same for ordinary light neutrino case as far as the top 10 list is concerned \cite{16iwata}.
Ten largest contributions (sum) cover 49.6$\%$ of the total contributions (sum) for the Fermi part, 52.3$\%$ for the Gamow-Teller part, and 13.4$\%$ for the tensor part. 
The minimum value for the tensor part is -0.0086 achieved by $(n~ l~ n'~ l')=(3~0~3~4)$ and $(3~4~3~0)$.

Correlation between the values of Eq.~(\ref{nupot4}) for different parts are examined in Fig.~\ref{fig3}. 
Comparison between Fermi and Gamow-Teller parts shows that they provide almost the same values, although the Fermi part generally shows slightly larger value compared to the Gamow-Teller part.
Such a quantitative similarity between Fermi and Gamow-Teller parts is not trivial since we can find essentially different mathematical representations at least in their form factors (cf.~Eq.~(\ref{eq-form})). 
The tensor part is positively correlated with the Fermi part (therefore Gamow-Teller part).
The $l=l'$ components of the tensor part contributions (sum) cover 28.9$\%$ of the total tensor part contributions (sum).

\section{Summary}
There are components of the two kinds in the nuclear matrix element; one is responsible for the amplitude and the other is for the cancellation.
As a component responsible for the amplitude, neutrino potential part (i.e., Eq.~(\ref{nupot})) is investigated in this article.
The presented results are valid not only to a specific double-beta decay candidates but also to all the possible candidates within $n, n' = 0, 1, \cdots, 3$ and  $l, l' = 0, 1, \cdots, 6$.
Note that, in terms of the magnitude, almost 40$\%$ smaller values are applied for the Gamow-Teller part in calculating the nuclear matrix element since $(g_V/g_A)^2 = (1/1.27)^2 \sim 0.62$ (cf. Eq.~(\ref{matrixel})).

Among several results on heavy neutrino cases, positive correlation of the values between Fermi, Gamow-Teller and tensor parts has been clarified.
This property is common to the light ordinary neutrino cases.
Apart from the tensor part values, almost a half of the total contributions has been shown to be occupied only by 10 largest contribution in which 10 largest contribution is exactly the same as $l =l'=0$ contributions.
As a result the enhanced dominance of $s$-wave contribution is noticed for heavy neutrino cases.

The other components of the NMEs also responsible for the cancellation will be studied in the next opportunity.

\end{document}